\documentclass[aps,pra,10pt,notitlepage,twocolumn]{revtex4-1}
\usepackage{graphicx}
\usepackage{amssymb,amsfonts}
\usepackage{amsmath}

\newcommand{\PRL}[1]{Phys. Rev. Lett.~\textbf{#1}}

\newcommand{\f}[1]{\textrm{#1}}
\newcommand{\x}{{\mathbf r}}
\newcommand{\K}{{\mathbf k}}
\newcommand{\dd}{{\mathrm d}}

\begin{document}

\author{T. Wasak$^1$, P. Sza\'nkowski$^1$, R. B\"ucker$^2$, J. Chwede\'nczuk$^1$ and M. Trippenbach$^1$}
\affiliation{$^1$Faculty of Physics, University of Warsaw, ul.\ Ho\.{z}a 69, PL--00--681 Warszawa, Poland\\
  $^2$Vienna Center for Quantum Science and Technology, Atominstitut, TU Wien, Stadionallee 2, 1020 Vienna, Austria}

\title{Bogoliubov theory for atom scattering into separate regions}

\begin{abstract}
  We review the Bogoliubov theory in the context of recent experiments, where atoms are scattered from a Bose-Einstein Condensate into two well-separated regions.
  We find the full dynamics of the pair-production process, calculate the first and second order correlation functions and show that the system is ideally
  number-squeezed. We calculate the Fisher information to show how the entanglement between the atoms from the two regions changes in time.
  We also provide a simple expression for the lower bound of the useful entanglement in the
  system in terms of the average number of scattered atoms and the number of modes they occupy.
  We then apply our theory to a recent ``twin-beam'' experiment [R. B\"ucker {\it et al.}, Nat. Phys. {\bf 7}, 608 (2011)].
  The only numerical step of our semi-analytical description can be easily solved and does not require implementation of any stochastic methods.
\end{abstract}
\maketitle

\section{Introduction}

In recent years, systems where strong correlations between particles are induced by pair-wise scattering, have attracted much attention.
In the canonical example, which is the parametric down-conversion, photon pairs are generated during the propagation of a laser beam through a non-linear medium.
The outcoming pairs of photons are entangled, and can serve as a probe of fundamental properties of quantum mechanics \cite{kwiat,horodecki},
%which is a clear manifestation of laws of quantum mechanics\cite{kwiat,horodecki}.
such as the Einstein-Podolsky-Rosen paradox, or violation of the Bell inequalities \cite{kwiat,horodecki,Reid09}.
On the other hand, entanglement can be exploited in practical applications, such as teleportation \cite{wootters1,wootters2}
or metrology beyond the Shot-Noise Limit (SNL) \cite{giovanetti,pezze}.

In this latter context, recent experiments with entangled states of atoms were a major breakthrough \cite{esteve}.
In \cite{esteve,maus, berrada},
two-body interactions were utilized to prepare non-classical squeezed states of atoms trapped in a double-well potential, which implies presence of many-body entanglement
\cite{sor}.
A similar idea was exploited to generate squeezing in the internal \cite{riedel,gross3,app} degrees of freedom.
In \cite{leroux,schl}, squeezing of a large spin of a collection of two-level atoms was achieved, using an intense laser field interacting with particles trapped in an optical cavity.

Simultaneously, a substantial experimental effort was put in order to generate entangled pairs of atoms scattered out of a Bose-Einstein Condensate (BEC).
In \cite{Perrin07,Krachmalnicoff10,Jaskula10,Kheruntsyan12}, a collision of two BECs lead to weak scattering of correlated atomic pairs onto a three-dimensional sphere
of initially unoccupied modes.
Although moderate number-squeezing between the opposite regions of the halo, and the related violation of the Cauchy-Schwarz inequality were experimentally demonstrated \cite{Jaskula10,Kheruntsyan12}, 
entanglement was never directly observed.
Alternatively, pair-production schemes were developed, where only few modes are strongly populated in a stimulated process, making the system somewhat easier to handle.
Stimulated four-wave-mixing processes have been implemented using different spin states of atoms \cite{dall,pertot,rugway} or Bragg scattering \cite{Vogels02,deng}.
Also, dynamic instabilities in moving optical lattices, populating modes with opposite quasi-momenta, have been used \cite{lattice,campbell}.
In \cite{bucker}, a BEC was transferred into the first excited state of a trapping potential and subsequent two-body collisions created a ``twin-beam'' system,
where stronger-than-classical correlations could directly be observed.

Analogous schemes have been implemented in internal atomic states, building upon spin-changing collisions
\cite{gross2, bookjans,smerzi}. Furthermore, in \cite{smerzi} it was shown that particles scattered in this process into a pair of $m_F=\pm1$ Zeeman sub-levels
are usefully entangled from the metrological point of view.

In this work we develop a theoretical model for the generic type of experiments, where particles scatter in pairs into two well-separated regions. If these regions are separated in the momentum
space, they could also be set apart by a sufficiently long expansion of the cloud. On the other hand, in cases when the regions are defined as two Zeeman sub-levels, 
they can be separated in space using a Stern-Gerlach scheme. Our model applies to any such possible configuration.
Therefore, the general conclusions of this study, concerning the correlations,
number-squeezing and entanglement, are valid for recent experiments \cite{dall,pertot,rugway,bucker,smerzi,lattice} in the regime,
where the depletion of the source BEC is low. We derive the Bogoliubov equations governing the dynamics of pair formation, and applying the Bloch-Messiah reduction \cite{braun2,sacha,single}, we
write the state in terms of pairs of independently squeezed modes.
We calculate the density and the number of scattered atoms, and the two body correlation between them.
We demonstrate the presence of ideal number-squeezing between the opposite regions, violation of the Cauchy-Schwarz inequality and,
using the Fisher information criterion known from quantum metrology, we show that the atoms from the twin-beam system are entangled.
We also provide a simple yet useful lower bound for the Fisher information in terms of the average number of scattered atoms, and the number of modes they occupy.
Finally, we apply the above formalism to the twin-beam experiment of \cite{bucker}.

This paper is organized as follows. In Sec.~\ref{universal}
we discuss the general properties of the solutions of the Bogoliubov equations. These observation allow to easily calculate the density and the
second order coherence of the system in Sec.~\ref{density} and the fluctuations of the population imbalance between the opposite regions in Sec.~\ref{squeezing}. In Sec.~\ref{sec_csi}
we take the first step towards the demonstration of particle entanglement present in the system, by showing that the second order correlation function violates the Cauchy-Schwarz inequality.
In Sec.~\ref{entanglement} we demonstrate that the scattered atoms are usefully entangled from the metrological point of view.
In Sec.~\ref{outline} we briefly describe the experimental setup of \cite{bucker} and in \ref{theory} derive the corresponding effective Bogoliubov equations.
Finally, in Sec.~\ref{results} we review the most relevant properties of the twin-beam system by showing the results of the numerical simulation.
We conclude in Sec.~\ref{conclusions}. %Some details of the analytical calculations and the numerical procedure are presented in the Appendices.

\section{General properties of the scattered particles}
\label{general}

We first present the general properties of the solution of the Bogoliubov equation, in cases where particles scatter pair-wise into well-separated regions.

\subsection{Bogoliubov equation for pair scattering}
\label{universal}

Our theoretical description of the pair-production process starts with a many-body Hamiltonian with contact two-body interactions
\begin{subequations}\label{ham}
  \begin{eqnarray}
    \hat H&=&\int\!\! d\x\,\hat\Psi^\dagger(\x)\left(-\frac{\hbar^2\nabla^2}{2m}+V(\x)\right)\hat\Psi(\x)\\
    &+&\frac g2\int\!\! d\x\,\hat\Psi^\dagger(\x)\hat\Psi^\dagger(\x)\hat\Psi(\x)\hat\Psi(\x).
  \end{eqnarray}
\end{subequations}
Here $V(\x)$ is an external trapping potential and $g=\frac{4\pi\hbar^2a}m$ is the strength of the two-body interactions, $a$ is the scattering length, $m$ is the atomic mass and
the field operator $\hat\Psi(\x)$ satisfies the bosonic commutation relations.
To derive the Bogoliubov equation, we first find the c-number (mean field) wave function of the BEC using the Gross-Pitaevskii Equation (GPE)
\begin{equation}
  i\hbar\partial_t\psi(\x)=\left(-\frac{\hbar^2\nabla^2}{2m}+V(\x)+g|\psi(\x)|^2\right)\psi(\x).
\end{equation}
We then write the field operator as a sum of the c-number part and the Bogoliubov correction, $\hat\Psi(\x)=\psi(\x)+\hat\delta(\x)$ and insert this expression into (\ref{ham}).
By keeping only the terms up to quadratic in $\hat\delta$ we obtain the Bogoliubov Hamiltonian
\begin{eqnarray}
  \hat H_{\rm bog}&=&\int\!\! d\x\,\hat\delta^\dagger(\x)\left(-\frac{\hbar^2\nabla^2}{2m}+V(\x)+2g|\psi(\x)|^2\right)\hat\delta(\x)\nonumber\\
  &+&g\int\!\! d\x\left(\hat\delta^\dagger(\x)\hat\delta^\dagger(\x)\,\psi^2(\x)+{\rm h.c.}\right).
\label{bog_ham}
\end{eqnarray}
The resulting Bogoliubov equation of motion is linear
\begin{eqnarray}
  i\hbar \partial_t \hat \delta(\x,t)&=&\left(-\frac{\hbar^2 \nabla^2}{2m} + V(\x) + 2 g |\psi(\x)|^2 \right) \hat \delta(\x,t)\nonumber\\
  &+&g \psi^2(\x) \hat \delta^\dagger(\x,t).\label{bog}
\end{eqnarray}
Usually, a numerical solution of this equation is found in a following way. The field operator is expanded in a basis of wave-functions $\tilde\varphi_i(\x)$ which match the geometry of the scattering problem
\begin{equation}\label{op_nondiag}
  \hat\delta(\x,t)=\sum_i\tilde\varphi_i(\x)\hat a_i(t).
\end{equation}
This expression is inserted into Eq.~(\ref{bog}), the resulting equation is multiplied by $\tilde\varphi^*_j(\x)$ and the outcome is integrated by sides over the whole space.
In effect, what we obtain is an equation of motion, which, through the matrices $\hat{\mathcal{A}}$ and $\hat{\mathcal{B}}$,  couples the evolution of the $j$-th
operator $\hat a_j(t)$, with (in general) all others operators
\begin{equation}\label{eq_mot}
  i\partial_t\hat a_j(t)=\sum_k\mathcal{A}_{jk}\hat a_k(t) + \sum_k\mathcal{B}_{jk}\hat a_k^{\dagger}(t).
\end{equation}
This equation is linear -- a consequence of the linearity of the Bogoliubov equation (\ref{bog}) -- so the general solution of (\ref{eq_mot}) reads
\begin{equation}\label{nondiag}
  \hat a_i(t)=\sum_j\mathcal{C}_{ij}(t)\hat a_j(0)+\sum_j\mathcal{S}_{ij}(t)\hat a_j^\dagger(0),
\end{equation}
where the matrices $\hat{\mathcal{C}}$ and $\hat{\mathcal{S}}$ satisfy
%$\hat{\mathcal{C}}\hat{\mathcal{C}}^\dagger-\hat{\mathcal{S}}\hat{\mathcal{S}}^\dagger=\hat{\mathbb{1}}$
$\hat{\mathcal{C}}\hat{\mathcal{C}}^\dagger-\hat{\mathcal{S}}\hat{\mathcal{S}}^\dagger=\hat{1}$ and $\hat{\mathcal{C}}\hat{\mathcal{S}}^{\rm T}-\hat{\mathcal{S}}\hat{\mathcal{C}}^{\rm T}=0$.
Later, we will apply this method to solve the Bogoliubov dynamics of the twin-beam production.
However, we will show in the following, that in cases where the detailed form of the Hamiltonian (\ref{bog_ham}) drives the scattering of atomic pairs into opposite regions
(as indeed happens in twin-beam experiments), the basic properties of the system can be deduced analytically if an appropriate set of mode functions $\varphi_i(\x)$ is chosen.

Let us denote the two separate regions into which the particles are scattered by L (left) and R (right). Particles populate L and R in a process of elastic scattering,
so the regions are usually separated in momentum space.
From this point of view, it is convenient to switch to the space of wave-vectors $\K$ and decompose the field operator as follows
\begin{equation}\label{op_diag}
  \hat\delta(\K,t)=\sum_i\varphi_{\rm R}^{(i)}\!(\K,t)\,\hat a^{(i)}_{\rm R}(t)+\sum_i\varphi_{\rm L}^{(i)}\!(\K,t)\,\hat a^{(i)}_{\rm L}(t).
\end{equation}
The operators $\hat a^{(i)}_{\rm R/L}(t)$ annihilate a particle in a mode characterized by the time-dependent wave function $\varphi_{\rm R/L}^{(i)}\!(\K,t)$,
which is localized in the right/left region in momentum space.
We underline, that this kind of separation is also present in position space after expansion of the cloud, or after application of a Stern-Gerlach pulse in internal-state experiments, respectively.
Moreover, the vector $\K$ might denote the quasi-momentum, if the scattering takes place in an optical lattice.

Formally, the only difference between the formulation (\ref{op_nondiag}) and (\ref{op_diag}) is the splitting of the field operator into the R and L modes.
However, for a linear equation of motion such as (\ref{bog}), there exists a unique basis of mode functions for which
the evolution equations of the mode pairs decouple from each other:
\begin{subequations}\label{pair_evolution}
  \begin{eqnarray}
    &&\hat a^{(i)}_{\rm R}(t)=c_i(t)\,\hat a^{(i)}_{\rm R}(0)+s_i(t)\,\hat a^{(i)\dagger}_{\rm L}(0)\\
    &&\hat a^{(i)}_{\rm L}(t)=c_i(t)\,\hat a^{(i)}_{\rm L}(0)+s_i(t)\,\hat a^{(i)\dagger}_{\rm R}(0),
  \end{eqnarray}
\end{subequations}
where $|c_i(t)|^2-|s_i(t)|^2=1$. This form of the Bogoliubov equation has a simple physical interpretation: atoms scatter pair-wise into opposite regions, and the total field operator (\ref{op_diag})
is a sum of independent mode pairs, which are squeezed in their relative population fluctuations, as will be explained in detail below.

Although the diagonal form (\ref{pair_evolution}) is much clearer than  (\ref{nondiag}), it is not obvious at the moment how this particular basis (\ref{op_diag}) can be found.
This is done in two steps, applying the procedure of the Bloch-Messiah reduction \cite{braun2,sacha,single}.
First, using Equations (\ref{op_diag}) and (\ref{pair_evolution}), we evaluate the one-body density matrix (first-order correlation function) and obtain
\begin{eqnarray}
  &&G^{(1)}(\K_1,\K_2;t)\equiv\langle\hat\delta^\dagger(\K_1,t)\hat\delta(\K_2,t)\rangle=\label{den_mat}\\
  &&=\sum_in_i\left(\varphi_{\rm R}^{(i)*}\!(\K_1,t)\varphi_{\rm R}^{(i)}\!(\K_2,t)+\varphi_{\rm L}^{(i)*}\!(\K_1,t)\varphi_{\rm L}^{(i)}\!(\K_2,t)\right),\nonumber
\end{eqnarray}
where $n_i=|s_i(t)|^2$. Note that $n_i$ is the average occupation of the $i$-th eigen-mode, and that a pair of modes $\varphi_{\rm R}^{(i)}\!(\K,t)$ and $\varphi_{\rm L}^{(i)}\!(\K,t)$ is degenerate
(has the same eigen-value $n_i$) due to the assumed symmetry between the left and the right region.
Since we are using the Heisenberg picture, the average value in Eq.~(\ref{den_mat}) and all equations that follow are calculated in the initial vacuum state of scattered atoms.

In any practical approach, if the basis (\ref{op_diag}) is not known a priori, this step first requires
a numerical evaluation of the density matrix (\ref{den_mat}) in any convenient basis (\ref{op_nondiag}), and subsequent diagonalization. Once this is done,
then according to Eq.~(\ref{den_mat}),
the basis functions $\varphi_{\rm R/L}^{(i)}\!(\K,t)$ are the momentary eigen-functions of the one-body density matrix (natural orbitals).
However, a second step is necessary to fully determine the functions $\varphi_{\rm R/L}^{(i)}\!(\K,t)$, because the density matrix -- contrary to the field operator (\ref{op_diag}) --
is insensitive to global phases of the mode functions. To retrieve this additional information, we calculate the anomalous density
\begin{eqnarray}
  &&M(\K_1,\K_2;t)\equiv\langle\hat\delta(\K_1,t)\hat\delta(\K_2,t)\rangle=\sum_i\sqrt{n_i(n_i+1)}\times\nonumber\\
  &&\times\left(\varphi_{\rm R}^{(i)}\!(\K_1,t)\varphi_{\rm L}^{(i)}\!(\K_2,t)+\varphi_{\rm R}^{(i)}\!(\K_2,t)\varphi_{\rm L}^{(i)}\!(\K_1,t)\right),\label{anom}
\end{eqnarray}
multiply it by sides with the eigen-functions of the density matrix and integrate over space. As a result, we retrive the information about the phases and obtain the full form of the mode functions
$\varphi_{\rm R/L}^{(i)}\!(\K,t)$ of the diagonal basis.

To summarize, we have outlined the structure of the solution of the Bogoliubov equation for cases where atoms are scattered into two opposite regions.
We will now show that the extra step, which is the transition from the ``numerical approach'' (\ref{op_nondiag}) to the diagonal basis (\ref{op_diag}), allows to easily determine the basic properties of the system of scattered atoms, like its density or higher correlation functions.

\subsection{Density and correlations}
\label{density}

The simplest observable characterizing the pair-production process is the density
\begin{equation}\label{den}
  \rho(\K;t)\equiv G^{(1)}(\K,\K;t)=\sum_in_i\left(\big{|}\varphi_{\rm R}^{(i)}\!(\K,t)\big{|}^2+\big{|}\varphi_{\rm L}^{(i)}\!(\K,t)\big{|}^2\right)
\end{equation}
which is, consistently with our derivation, localized in the two opposite regions. By integrating the above function over space, we obtain the information about the expected number of
scattered atoms as a function of time
\begin{equation}\label{nscat}
  \left\langle\hat N\right\rangle=\int\!\dd\K\,\rho(\K;t)=2\sum_in_i.
\end{equation}
Additional information about the system is carried by the correlations between the scattered particles.
The probability of simultaneous detection of two atoms at momenta $\K_1$ and $\K_2$ can be obtained from the normalized second-order correlation function
\begin{equation}
  g^{(2)}(\K_1,\K_2;t)=\frac{\langle\hat\delta^\dagger(\K_1,t)\hat\delta^\dagger(\K_2,t)\hat\delta(\K_2,t)\hat\delta(\K_1,t)\rangle}{\rho(\K_1;t)\rho(\K_2;t)}.\label{g2}
\end{equation}
According to the Wick's theorem, this function can be written in terms of the one-body density matrix (\ref{den_mat}) and the anomalous density (\ref{anom}) as follows
\begin{equation}
  g^{(2)}(\K_1,\K_2;t)=1+\frac{|G^{(1)}(\K_1,\K_2;t)|^2+|M(\K_1,\K_2;t)|^2}{\rho(\K_1;t)\rho(\K_2;t)}.\label{g2_w}
\end{equation}
The transition from Eq.~(\ref{g2}) to (\ref{g2_w}) might seem an unnecessary complication, however we will argue that it allows
for a simple and intuitive interpretation of the second-order correlation function. % Namely, we now exploit the information provided by the diagonal basis of the Bogoliubov equation.
According to Eq.~(\ref{den_mat}), the density matrix is non-vanishing only when $\K_1$ and $\K_2$ are both either in the right or left region, so $|G^{(1)}|^2$  governs the Hanbury-Brown and Twiss (HBT) type of
local correlations. On the other hand, as can be seen from Eq.~(\ref{anom}), the anomalous density is non-zero only when $\K_1$ and $\K_2$
are in the opposite regions,
so it describes the cross-correlations between the two members of the scattered pair.
Clearly,
this simple interpretation of the second order correlation function as a sum of local- and opposite-momentum correlations would
have been much more difficult if we had not applied the diagonalization procedure and the Wick's theorem.

%Assuming a symmetric system, the correlation coefficient $C = \mathcal G^{(2)}_\mathrm{RL}/( \mathcal G^{(2)}_\mathrm{RR} \mathcal G^{(2)}_\mathrm{LL})^{1/2}$ can now be expressed as
%\begin{equation}
%C = \frac{ \int \mathrm{d}\K_1 \int \mathrm{d}\K_2 |M(\K_1,\K_2;t)|^2 + (\langle\hat N \rangle/2)^2} {\int \mathrm{d}\K_1 \int \mathrm{d}\K_2 |G^{(1)}(\K_1,\K_2;t)|^2  + (\langle\hat N \rangle/2)^2},
%\end{equation}
%where all integrations run over the entire momentum space.
%Stronger-than-classical correlations are witnessed by $C > 1$.

\subsection{Number squeezing}
\label{squeezing}

Another property characterizing the scattering process are the fluctuations of the population imbalance between the two regions.
If these fluctuations are suppressed below the properly defined shot-noise level, the system is {\it number squeezed},
which proves that the atoms scatter in pairs rather then independently to the left and to the right region. A quantitative description of the number squeezing
involves the left and right atom number operators defined as the integrals of the density operators over the corresponding volumes, i.e.
\begin{equation}\label{at_num}
  \hat N_{\rm R/L}=\int\limits_{\rm R/L}\!\!\dd\K\,\hat\delta^\dagger(\K,t)\hat\delta(\K,t).\\
\end{equation}
The population imbalance operator is then simply defined as $\hat n=\hat N_{\rm R}-\hat N_{\rm L}$ and using Eq.~(\ref{op_diag}) we obtain
\begin{equation}\label{pop_imb}
  \hat n=\sum_i\left(\hat a_{\rm R}^{(i)\dagger}(0)\,\hat a_{\rm R}^{(i)}(0)-\hat a_{\rm L}^{(i)\dagger}(0)\,\hat a_{\rm L}^{(i)}(0)\right).
\end{equation}
The number squeezing factor is defined as
\begin{equation}
  \xi^2=\frac{\Delta^2\hat n}{\left\langle\hat N\right\rangle},
\end{equation}
where $\Delta^2\hat n=\langle\hat n^2\rangle-\langle\hat n\rangle^2$ is the variance of the population imbalance operator. If the fluctuations between the two regions are suppressed below the shot-noise
level defined as $\xi^2=1$,
the system is called ``number-squeezed''.
In our case, since $\hat n$ does not depend on time and the initial state is a vacuum, we obtain that $\xi^2\equiv0$. Therefore, the two-region Bogoliubov system
is perfectly number-squeezed, as anticipated in the previous section.

The ideal number-squeezing is a result of clear separation of the two scattering regions. In such case, it is natural to define the local atom-number operators (\ref{at_num}) and the population
imbalance operator (\ref{pop_imb}). It is important to note that not all systems, where particles are scattered in pairs are perfectly number squeezed.
For instance, when two Bose-Einstein condensates collide, they produce a halo of atoms due to two-body elastic scattering into the initially unoccupied modes \cite{ketterle,perrin}.
In this system however, there is no simple way to define two separate regions. One can instead measure the number of atoms in two bins lying on the opposite sides of the halo.
Moderate number-squeezing of the atom number difference between these bins has been observed experimentally \cite{squeezing}, but it is impossible to reach the limit $\xi^2=0$ \cite{wasak}.
In contrast, the twin matter wave configurations \cite{bucker,dall,rugway,lattice, smerzi},
are ideal sources of correlated atomic pairs occupying two well-defined areas.

\subsection{Violation of the Cauchy-Schwarz inequality}
\label{sec_csi}

Apart from the number squeezing, the twin-region system can be characterized by another expression, which is called the Cauchy-Schwarz inequality.
It relates the strength of the local and opposite correlations to witness the pair-scattering process. Following \cite{Kheruntsyan12}, we define averaged second-order correlations as
%\begin{eqnarray}
%\mathcal{G}^{(2)}_{\mu \nu} &\equiv& \int_\mu \mathrm{d}\K_1 \int_\nu \mathrm{d}\K_2 \langle\hat\delta^\dagger(\K_1,t)\hat\delta^\dagger(\K_2,t) \hat\delta(\K_2,t)\hat\delta(\K_1,t)\rangle \nonumber \\
%&=& \langle \hat N_\mu \hat N_\nu \rangle + \delta_{\mu \nu} \langle \hat N_\mu \rangle,
%\end{eqnarray}
\begin{equation}
  \mathcal{G}^{(2)}_{\mu \nu} \equiv \int\limits_\mu \mathrm{d}\K_1 \int\limits_\nu \mathrm{d}\K_2 \langle\hat\delta^\dagger(\K_1,t)\hat\delta^\dagger(\K_2,t) \hat\delta(\K_2,t)\hat\delta(\K_1,t)\rangle
  \label{eq:summed_g2}
\end{equation}
where $\mu,\nu \in \{\mathrm{R},\mathrm{L}\}$.
In the symmetric case, the Cauchy-Schwarz inequality $\mathcal G^{(2)}_\mathrm{RL} \leqslant ( \mathcal G^{(2)}_\mathrm{RR} \mathcal G^{(2)}_\mathrm{LL})^{1/2}$ can now be re-written as
\begin{equation}
  \int\limits_{\rm R}\!\!\mathrm{d}\K_1\!\!\int\limits_{\rm L}\!\!\mathrm{d}\K_2 |M(\K_1,\K_2;t)|^2 \leqslant \int\limits_{\rm R}\!\!\mathrm{d}\K_1\!\!\int\limits_{\rm R}\!\!\mathrm{d}\K_2|G^{(1)}(\K_1,\K_2;t)|^2.
  \label{CSI_one_body}
\end{equation}
Using expressions (\ref{den_mat}) and (\ref{anom}) we obtain
\begin{subequations}
  \begin{eqnarray}
    &&\mathcal{G}^{(2)}_\mathrm{RL}=\left(\sum_in_i\right)^2+\sum_in_i(n_i+1)\\
    &&\mathcal{G}^{(2)}_\mathrm{RR}=\left(\sum_in_i\right)^2+\sum_in_i^2
  \end{eqnarray}
\end{subequations}
thus the Cauchy-Schwarz inequality reads
\begin{equation}
  \sum_in_i(n_i+1)\leqslant\sum_in_i^2,
  \label{eq:csi_violation}
\end{equation}
which is true only for all $n_i=0$.
As soon as particles start to scatter into the two regions, the Cauchy-Schwarz inequality is clearly violated. To quantify the degree of violation, a coefficient $\mathcal{C}$ was introduced in
\cite{Kheruntsyan12}, which reads
\begin{equation}
  \mathcal{C}=\frac{\mathcal{G}^{(2)}_\mathrm{RL}}{\mathcal{G}^{(2)}_\mathrm{RR}}.
\end{equation}
When $\mathcal{C}\leqslant1$ the system is in the ``classical'' regime, while $\mathcal{C}>1$ signify correlations which are stronger than allowed by the classical physics.
In our case this coefficient reads
\begin{equation}\label{coeff_c}
  \mathcal{C}=1+\frac{\sum_in_i}{\left(\sum_in_i\right)^2+\sum_in_i^2}.
\end{equation}
Clearly, always $\mathcal{C}>1$, because it is a sum of unity and a non-negative part.
For high mode populations $n_i$, the second term, which is inversely proportional to the number of scattered particles tends to zero, restoring the classical limit. Nevertheless,
as demonstrated with photons in \cite{marino},
the confidence by which the Cauchy-Schwarz inequality can be violated in the presence of classical noise still \emph{increases} with more strongly populated modes.
However, it is the Fisher information, which is the quantity highly sensitive to particle entanglement in the high-gain regime, as we show in the following section.
This measure quantifies the potential for sub-shot-noise interferometry, and increases with rising mode population,
in spite of the decreasing ``granularity'' of the matter wave \cite{single} that leads to all second-order correlation functions approaching equal values.

\subsection{Entanglement and interferometry}
\label{entanglement}

We now show that atoms occupying the two regions are entangled, and could be used as an input of a quantum interferometer operating below the shot-noise level.
We first recall how the precision of the phase estimation is related to the entanglement of input states using as an example the standard two-mode Mach-Zehnder Interferometer (MZI).
Then, we extend these concepts to the case, where the interferometer operates between two regions, each having a multi-mode structure determined by the Bogoliubov equations.

When speaking about two-mode interferometers, it is convenient to introduce a set of three operators
\begin{subequations}\label{ang}
  \begin{eqnarray}
    \hat J_x&=&\frac12\left(\hat a^\dagger_{\rm R}\hat a^{\phantom{\dagger}}_{\rm L}+\hat a^\dagger_{\rm L}\hat a^{\phantom{\dagger}}_{\rm R}\right)\\
    \hat J_y&=&\frac1{2i}\left(\hat a^\dagger_{\rm R}\hat a^{\phantom{\dagger}}_{\rm L}-\hat a^\dagger_{\rm L}\hat a^{\phantom{\dagger}}_{\rm R}\right)\\
    \hat J_z&=&\frac12\left(\hat a^\dagger_{\rm R}\hat a^{\phantom{\dagger}}_{\rm R}-\hat a^\dagger_{\rm L}\hat a^{\phantom{\dagger}}_{\rm L}\right)
  \end{eqnarray}
\end{subequations}
which obey the same commutation relations as the angular momentum operators.
The MZI, which is an interferometric device, where the imprint of the phase $\theta$ onto the input state is preceded and followed by a pair of symmetric beam-splitters,
can be represented by a unitary evolution
operator $\hat U(\theta)=e^{-i\theta\hat J_y}$. If the phase is estimated in a series of $\nu\gg1$ measurements performed on the output state,
the precision of the phase estimation is limited by the Cramer-Rao Lower Bound (CRLB) \cite{helstrom,cramer},
\begin{equation}
  \Delta\theta\geqslant\frac1{\sqrt\nu}\frac1{\sqrt F_{\rm Q}}.\label{crlb}
\end{equation}
Here, $F_Q$ is the Quantum Fisher Information (QFI), which is related to the unitary transformation $\hat U(\theta)$. For pure states transformed by the MZI it is equal to $F_{\rm Q}=4\,\Delta^2\!\hat J_y$,
where the variance is calculated in the input state of the interferometer \cite{braun}. The CRLB states, that if $\theta$ is determined using any possible type of measurement and estimator, then
 the precision $\Delta\theta$ is bounded as in Eq.~(\ref{crlb}).

Apart from providing a lower bound for the error of the phase estimation, the $F_{\rm Q}$ is an entanglement measure. Namely, when the input state has an average number of $\langle\hat N\rangle$ particles, then
if $F_{\rm Q}>\langle\hat N\rangle$, the state is particle-entangled \cite{sorensen,giovanetti,pezze,hyllus}.

We now show, that a natural extension of the two-mode picture allows to employ the concept of the QFI as an entanglement measure also in our multi-mode system of interest.
To this end, we introduce the following analog of the two-mode angular momentum operators (\ref{ang}),
\begin{subequations}\label{ang_twin}
  \begin{eqnarray}
    \hat J_x&=&\frac12\int\limits_{\rm R}\!\!\dd\K\,\left(\hat\delta^\dagger(\K)\hat\delta(-\K)+\hat\delta^\dagger(-\K)\hat\delta(\K)\right)\\
    \hat J_y&=&\frac1{2i}\int\limits_{\rm R}\!\!\dd\K\,\left(\hat\delta^\dagger(\K)\hat\delta(-\K)-\hat\delta^\dagger(-\K)\hat\delta(\K)\right)\\
    \hat J_z&=&\frac12\int\limits_{\rm R}\!\!\dd\K\,\left(\hat\delta^\dagger(\K)\hat\delta(\K)-\hat\delta^\dagger(-\K)\hat\delta(-\K)\right),
  \end{eqnarray}
\end{subequations}
where we dropped the explicit time-dependence of the $\hat\delta(\K,t)$ to simplify the notation. Also, for simplicity, we choose the well-separated
regions R and L to be localized symmetrically on the opposite sites of $\K=0$. The construction of these operators, which satisfy the same commutation relations as (\ref{ang}), is
based on the analogy between the two-mode systems and the twin-beam configuration. In the former case, the operators connect the right and left modes, while in the latter the left
and right sub-spaces. Such a definition (\ref{ang_twin}) is meaningful only in situations, where the system consists of two well-separated sub-systems.%, as is the case of the twin-beam cofniguration.

Using the decomposition of the field operator into the set of independent modes, Equations (\ref{op_diag}) and (\ref{pair_evolution}), the above integrals yield, that
each angular momentum operator is a sum of operators acting on each mode independently, that is
\begin{subequations}\label{ang2}
  \begin{eqnarray}
    \hat J_x&=&\sum_i\frac12\left(\hat a^{(i)\dagger}_{\rm R}\hat a^{(i)}_{\rm L}+\hat a^{(i)\dagger}_{\rm L}\hat a^{(i)}_{\rm R}\right)\equiv\sum_i\hat J_x^{(i)}\\
    \hat J_y&=&\sum_i\frac1{2i}\left(\hat a^{(i)\dagger}_{\rm R}\hat a^{(i)}_{\rm L}-\hat a^{(i)\dagger}_{\rm L}\hat a^{(i)}_{\rm R}\right)\equiv\sum_i\hat J_y^{(i)}\\
    \hat J_z&=&\sum_i\frac1{2}\left(\hat a^{(i)\dagger}_{\rm R}\hat a^{(i)}_{\rm R}-\hat a^{(i)\dagger}_{\rm L}\hat a^{(i)}_{\rm L}\right)\equiv\sum_i\hat J_z^{(i)}.
  \end{eqnarray}
\end{subequations}
These expressions show again that it is natural to describe the two-region system using the diagonal basis (\ref{op_diag}). In this language, the angular momentum operators are simply a sum of
operators acting on each pair of modes independently, which vastly simplifies the further analysis.

To establish a direct relation between the two-mode and two-region case, we now assume that the system is transformed in the multi-mode analog of the Mach-Zehnder interferometer.
As outlined above, to demonstrate the presence of useful entanglement between the atoms in the left and in the right, it is necessary to calculate the QFI. Using Eq.~(\ref{ang2}) we obtain that
\begin{eqnarray}
  F_{\rm Q}&=&4\,\Delta^2\!\hat J_y=4\left\langle\left(\sum_i\hat J_y^{(i)}\right)^2\right\rangle\\
  &=&4\sum_i\left\langle\left(\hat J_y^{(i)}\right)^2\right\rangle+4\sum_{i\neq j}\left\langle\hat J_y^{(i)}\hat J_y^{(j)}\right\rangle.
\end{eqnarray}
Since the construction of the basis (\ref{op_diag}) explicitly assumes that each mode is independent from all others, the second term in the last equality is
$4\sum_{i\neq j}\left\langle\hat J_y^{(i)}\right\rangle\left\langle\hat J_y^{(j)}\right\rangle=0$, because the symmetry between the $R$ and $L$ regions implies that
$\left\langle\hat J_y^{(i)}\right\rangle=0$ for all $i$. Therefore we obtain that the QFI is equal to
\begin{equation}\label{qfi}
  F_{\rm Q}=4\sum_i\left\langle\left(\hat J_y^{(i)}\right)^2\right\rangle=4\sum_in_i^2+2\left\langle\hat N\right\rangle,
\end{equation}
where the last equality comes directly from the substitution of (\ref{pair_evolution}) into the definition of the $\hat J_y^{(i)}$ operator.
Also, we used $\langle\hat N\rangle=2\sum_in_i$, according to Eq.~(\ref{nscat}).
Clearly $F_{\rm Q}>\langle\hat N\rangle$, so the system is entangled. Moreover, one can refer the QFI to the ultimate bound for the precision of the parameter estimation, which is the Heisenberg
limit. For a system with fluctuating number of particles, this upper bound is equal to $\langle\hat N^2\rangle$. Using (\ref{pair_evolution}) again, we obtain, that \cite{hyllus}
\begin{eqnarray}
  \left\langle\hat N^2\right\rangle &=& \left\langle\left(\sum_i\left(\hat a^{(i)\dagger}_{\rm R}\hat a^{(i)}_{\rm R}+\hat a^{(i)\dagger}_{\rm L}\hat a^{(i)}_{\rm L}\right)\right)^2\right\rangle \\
  &=& 8\sum_in_i^2+\left\langle\hat N\right\rangle.
\end{eqnarray}
For a large number of scattered particles, when $\langle\hat N\rangle\ll\sum_in_i^2$, we obtain $F_{\rm Q}\simeq\frac12\langle\hat N^2\rangle$. The value of the QFI, which
is only one-half smaller than the Heisenberg Limit is a clear indication of very strong entanglement present in the system in the high-gain regime.
At intermediate times, $F_{\rm Q}<\frac12\langle\hat N^2\rangle$ due to mode competition, which has a negative impact on the entanglement as witnessed by the QFI \cite{szan}. To picture this,
consider a ``frustrated case'', where all atoms scatter uniformly into $M$ pairs of modes, so that all $n_i\equiv n$ are equal.
In this case, the number of scattered atoms is simply $\langle\hat N\rangle=2nM$, and the QFI is $F_{\rm Q}=4n^2M+2\left\langle\hat N\right\rangle$. The QFI normalized to the SNL is
\begin{equation}\label{SNL}
  \frac{F_{\rm Q}}{\left\langle\hat N\right\rangle}=2+\frac{\left\langle\hat N\right\rangle}{M}.
\end{equation}
When, on average, there is less than a particle per a set of modes, i.e. $\frac{\left\langle\hat N\right\rangle}{M}\ll1$,
the QFI surpasses the SNL only by a factor of 2, a natural reminiscence of atoms being scattered in pairs. Equation (\ref{SNL}) is a simple yet intuitive estimation of
the lower bound of useful entanglement in terms of the number of scattered atoms and occupied modes.

\section{Application: twin-beam system}

We now apply the above formalism to the twin-beam system of \cite{bucker}. First, we describe the physical mechanism which leads to the creation of the two correlated beams.
As shown below, some basic information about the dynamics of the pair production allow to construct a simple one-dimensional Bogoliubov model, which can be easily solved numerically.

\subsection{Scheme of the experiment}
\label{outline}

The experimental sequence applied in \cite{bucker} to produce correlated atom pairs was following.
First, an almost pure Bose-Einstein Condensate (BEC) of $N_0\approx 800$ $^{87}$Rb atoms with scattering length equal to $a = 5.3$ nm was created at temperature $T\approx 25$\,nK.
The cloud was trapped in an approximately harmonic potential
\begin{equation}
  V(\x) \simeq \frac12m\omega_x^2x^2+\frac12m\omega_y^2y^2+\frac12m\omega_z^2z^2,\label{pot}
\end{equation}
where atomic mass is equal to $m=1.44\times10^{-25}\ \f{kg}$, and the frequency $\omega_x=2\pi\times 16.3\ \f{Hz}$ is much smaller than $\omega_y=2\pi\times 1.83\ \f{kHz}$ and $\omega_z=2\pi\times 2.50\ \f{kHz}$, so the BEC is strongly elongated along the $x$-axis.

After the BEC was created, the trapping potential was shaken in a controlled way, so the atoms were transferred to the first excited state along the $y$-direction.
In order to achieve the maximal transfer efficiency, the shaking was optimized using quantum optimal control theory  \cite{bucker2}.
Afterwards, binary collisions transfered particle pairs to the ground state of the potential, and the excess energy $2\hbar\omega_y$ was converted into back-to-back movement of the two atoms along $x$.
Momentum conservation ensured, that their wave vectors had equal lengths
$k_0\approx\sqrt{2m\omega_y/\hbar}$ and point in opposite directions.
%This is the pair scattering process, which leads to production of two well defined counter-propagating matter-waves.
Small corrections to the value of $k_0$ may arise from an effective mean-field potential, as will be discussed below.

\subsection{Theoretical description}
\label{theory}

Neglecting thermal phase fluctuations along the elongated direction $x$, which is valid at very low temperatures only~\cite{petrov},
the condensate wave function acting as a source for the pair-production can be found by solving the stationary GPE
\begin{equation}\label{stat_gpe}
  \mu\,\psi(\x)=\left(-\frac{\hbar^2\nabla^2}{2m}+V(\x)+g|\psi(\x)|^2\right)\psi(\x),
\end{equation}
where $\mu$ is the chemical potential.
This function can be evaluated numerically, by referring to the description of the experiment from the previous section, and noting that after the shaking of the trap,
the BEC is in the first excited state $n_y=1$ along the $y$ axis and in the ground state $n_z=0$ along $z$. However, this can be approximated by an analytical expression, as argued below.

First note, that since the characteristic energies $\hbar\omega_y$ and $\hbar\omega_z$ are large, and the number of atoms in the BEC is small, the non-linear term can be safely neglected in evaluation of the eigenstates along $y$ and $z$. As a result, assuming that the total wave-function $\psi(\x)$ separates in three directions (which has been confirmed numerically), we obtain
\begin{equation}\label{bec}
  \psi(\x) = \phi(x)\times \psi^{\rm (ho)}_{n_y=1}(y)\times \psi^{\rm (ho)}_{n_z=0}(z),
\end{equation}
where the functions $\psi^{\rm (ho)}_{n_y=1}(y)$ and $\psi^{\rm (ho)}_{n_z=0}(z)$ are the eigen-states of the one-dimensional harmonic potential in $y$ and $z$ correspondingly.
The function $\phi(x)$ is found by inserting the above expression into Eq.~(\ref{stat_gpe})  and integrating out the orthogonal directions. As a result, we obtain an effective equation
\begin{equation}\label{gpex}
  (\mu-\epsilon_\perp)\, \phi(x) = \left( -\frac{\hbar^2}{2m}\frac{ \partial^2}{\partial x^2} + \tilde g |\phi(x)|^2 + \frac12 m \omega_x^2x^2\right)\phi(x)
\end{equation}
where zero-point energy equals $\epsilon_\perp = \frac32\omega_y + \frac12\omega_z$ and the non-linearity reads
\begin{eqnarray}
  \tilde g&=&g \left[\int\!\! \mathrm{d}y\, \left(\psi^{\rm (ho)}_{n_y=1}(y)\right)^4\right]\times \left[\int\!\! \mathrm{d}z\, \left(\psi^{\rm (ho)}_{n_z=0}(z)\right)^4\right]\nonumber\\
  &=&\frac{3}{8\pi}\frac g{a_{\mathrm{ho},y}\, a_{\mathrm{ho},z}}.
\end{eqnarray}
Here $a_{\mathrm{ho},i}=\sqrt{\frac{\hbar}{m\omega_i}}$ are the harmonic oscillator lengths for $i=y,\,z$.
Since the trap is shallow in the $x$ direction, the solution of the stationary GPE (\ref{gpex}) can be well approximated by the Thomas-Fermi (TF) formula \cite{dalfovo}
\begin{equation}
  \phi(x) =\sqrt{ \frac{\tilde\mu}{\tilde g}}\sqrt{1 - \frac{x^2}{R_{\rm tf}^2}}.
\end{equation}
where the effective chemical potential is
\begin{equation}
  \tilde \mu =\mu-\epsilon_\perp =\left( \frac{3 \tilde g N_0 \sqrt{m \omega_x^2 }}{ 4\sqrt{2} } \right)^{2/3}=492\,{\rm Hz}\times h,
\end{equation}
leading to a TF radius of $R_{\rm tf}=\sqrt\frac{2\tilde \mu}{m \omega_x^2}=20.75\,\mu{\rm m}$.

Within the approximation of neglecting thermal phase fluctuations, we have fully determined the wave-function of the BEC, which we insert into the Bogoliubov Hamiltonian (\ref{bog_ham}).
Next, we expand the field operator $\hat\delta(\x)$ in an orthonormal basis. Along the $y$ and $z$ directions, it is natural to use the eigen-states of the harmonic oscillator as the basis functions,
since it matches the geometry of the source BEC. Along the $x$ direction, we use a plane-wave basis, and get
\begin{equation}
  \hat \delta(\x,t) = \sum_{n_y,n_z}\int\!\frac{dk}{2\pi} e^{i k x} \psi^{\rm (ho)}_{n_y}(y)\psi^{\rm (ho)}_{n_z}(z) \hat \delta(k,n_y,n_z,t).
\end{equation}
Since the atom pairs are emitted into the ground state along $y$ only (which is ensured by the anisotropy and anharmonicity of the potential), the sum over the eigen-states can be safely truncated at $n_y=0$ and $n_z=0$. This reduces the dynamics of the pair-production to one-dimensional problem along the $x$ axis, with the orthogonal directions frozen out, i.e.
\begin{equation}\label{frozen}
  \hat \delta(\x,t) \simeq \psi^{\rm (ho)}_{n_y=0}(y)\psi^{\rm (ho)}_{n_z=0}(z)\sum_k \frac{e^{i k x}}{\sqrt L}\,  \hat a_k(t),
\end{equation}
where $L$ is the quantization volume.
We insert this field operator into Eq.~(\ref{bog_ham}), evaluate the spatial integrals and upon the change of variables $\hat a_k(t)e^{ i \tilde\mu t}\rightarrow\hat a_k(t)$ obtain
\begin{eqnarray}
  &&\hat H_{\rm bog}\simeq\sum_k\left( \frac{\hbar^2 k^2}{2m}-(\tilde\mu+\hbar\omega_y)\right)\hat a^\dagger_k(t)\hat a^{\phantom{\dagger}}_k(t)+\label{ham_fin}\\
  &&\sum_{k,k'}\left(2f_{k-k'}\,\hat a^\dagger_k(t)\hat a^{\phantom{\dagger}}_{k'}(t)+f_{k+k'}\left(\hat a^\dagger_k(t)\hat a^{{\dagger}}_{k'}(t)+\rm{h.c.}\right)\right),\nonumber
\end{eqnarray}
where $f(q)=\frac23\frac{\tilde g}L\!\int\!dx\, e^{-iqx} \phi^2(x)$. We solve the resulting Bogoliubov equation numerically, using a fourth order Runge-Kutta method, and find the matrices
$\hat{\mathcal{C}}$ and $\hat{\mathcal{S}}$ as defined in Eq.~(\ref{nondiag}).

Using the above Hamiltonian, one can also analytically determine $k_0$, i.e. the position of the central peak. To this end, we employ a two-mode approximation by replacing the function $f(q)$ with a Dirac delta, and obtain the Bogoliubov equation
\begin{equation}
  i \hbar \partial_t \hat \delta_k(t)=\frac{\hbar^2}{2m}\left(k^2-k_0^2\right)\hat \delta_k(t)+\frac23\,\tilde\mu\,\hat\delta^\dagger_{-k}(t),
\end{equation}
where $k_0$ is shifted with respect to the harmonic excitation energy due to the mean-field repulsion and reads
\begin{equation}\label{k0}
  k_0= \sqrt{\frac{2m}{\hbar^2}\left(\hbar\omega_y-\frac{\tilde\mu}3\right)}=5.35\,\mu{\rm m}^{-1}.
\end{equation}
%\todo{In the numerical result the peak is at 5.45. Is there any intuitive reason for this discrepancy (like e.g. the spatial inhomogeneity)? Doesn't really matter, actually...}
This result is in good agreement with the experimentally measured position of the peak density, i.e. $k_{0,{\rm exp}}=5.55\,\mu{\rm m}^{-1}$.

\subsection{Numerical results}
\label{results}

\begin{figure}[htb!]
  \includegraphics[clip, scale=0.4]{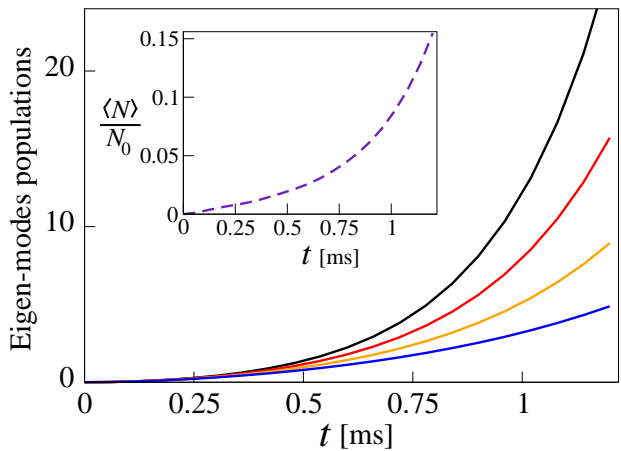}
  \caption{(color online) Populations of the first four eigen-modes of the density matrix (i.e. the eigen-values) as a function of time.
    The inset shows the average number of scattered atoms
    normalized to the number of atoms in the BEC as a function of time. The Bogoliubov approximation is valid for as long as this number is much smaller than one.
    In our case, we interrupt the calculations at $t=1.2$\,ms, when $\frac{\langle\hat N\rangle}{N_0}\simeq0.15$.}\label{fig_vals}
\end{figure}

In this section, we display the most important characteristics of the twin-beam system, starting from the solution of the eigen-problem of the density matrix (\ref{den_mat}).
In Fig.~\ref{fig_vals} we plot the first four eigen-values of the density matrix, as a function of time.
The inset shows the total number of scattered atoms $\langle\hat N\rangle$ normalized to the
occupation of the BEC, as a function of time. The Bogoliubov approximation is valid for as long as $\langle\hat N\rangle\ll N_0$, so we interrupt the simulation at $t=1.2$\,ms,
when $\langle\hat N\rangle\simeq15\%N_0$. For longer times, when the depletion of the BEC cannot be neglected, a atom-number conserving method, such as the one introduced in \cite{bucker3} must be used.

\begin{figure}[htb!]
  \includegraphics[clip, scale=0.5]{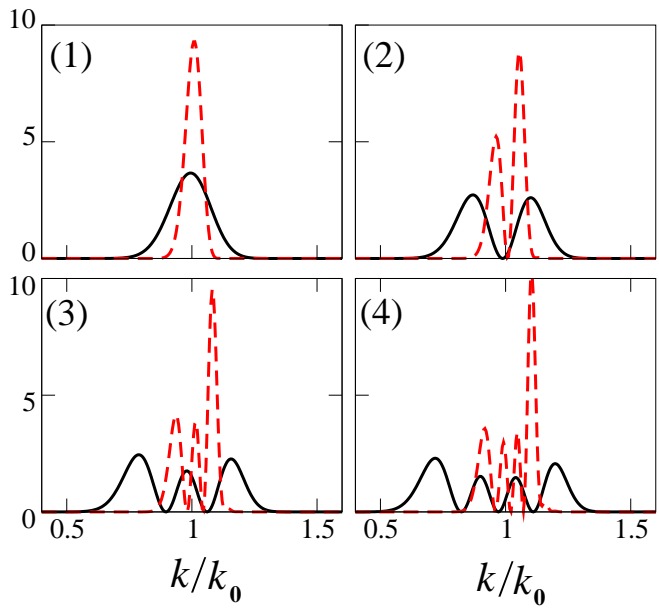}
  \caption{(color online) The modulus square of the first four eigen-vectors localized in the right sub-space, i.e. $|\varphi_{\rm R}^{(i)}\!(k)|^2$ with $i=1,2,3,4$.
    The solid black lines are results of diagonalization of the density-matrix at $t=0.1$\,ms while the dashed red lines at $t=1.2$\,ms. The figure shows how due to the time-energy uncertainty
    relation, the eigen-vectors narrow in the course of time around the central wave-vector $k_0$.
  }\label{fig_vecs}
\end{figure}

In Fig.~\ref{fig_vecs} we plot the first four eigen-vectors of $G^{(1)}$ localized in the right half-space,
i.e. $|\varphi_{\rm R}^{(i)}\!(k)|^2$ with $i=1,2,3,4$, calculated at an early time $t=0.1$\,ms and at $t=1.2$\,ms.
Due to the time-energy uncertainty relation, the eigen-modes localize around $k=k_0$ at later times.

\begin{figure}[htb]
 \includegraphics{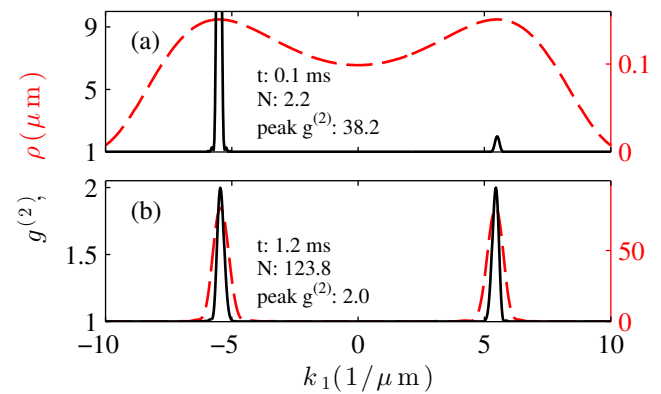}
  \caption{(color online) Normalized second-order correlation functions $g^{(2)}(k_1,k_2\equiv k_0;t)$ for fixed $k_2$ (solid lines, left $y$-axis),
    and density profiles $\rho(k_1; t)$ (dashed lines, right axis) in momentum space.
    The results are calculated at $t=0.1$\,ms (a) and  $t=1.2$\,ms (b).
    At early times, many momentum modes are occupied and the width of $g^{(2)}$ is much smaller than the beam size.
    Later, two distinct peaks emerge, which are almost single-mode.
  }\label{fig_den}
\end{figure}

This can be seen even more clearly, by plotting the density $\rho(k;t)$ at these two instants, as shown in Fig.~\ref{fig_den} (dashed lines). At $t=0.1$\,ms,
two broad beams start to form on top of the uniform density. Later, at $t=1.2$\,ms, strongly localized peaks clearly dominate over the flat background.
On top of these curves, we plot the normalized second-order correlation function as defined in
Eq.~(\ref{g2}), with one of the arguments set equal to the resonant wave-vector $k_0$, i.e. $g^{(2)}(k_1,k_2\equiv k_0;t)$. At $t=0.1$\,ms, the cross-correlation,
which is governed by the anomalous density, is very large, i.e. $g^{(2)}(-k_0,k_0;0.1\,{\rm ms})\simeq40$. This is a characteristic property of the Bogoliubov system in the low-occupation
regime \cite{wasak}, and indicates strong violation of the Cauchy-Schwarz inequality~(\ref{CSI_one_body}). Also, for this early time, the width of both $g^{(2)}$ peaks are much more narrow than the beam size.
This is consistent with the results shown in Fig.~\ref{fig_vals}, where at early times many eigen-mode pairs of the density matrix are almost equally occupied.
%The narrowing of the $g^{(2)}$ peaks with respect to the density is a result of random shot-to-shot interfetences between the independent modes.
At later times, when a single pair of modes start to become dominant, the width of the peak in $g^{(2)}$ and the system size approach each other.
While this corresponds to beams that are single-mode with respect to their local one-body properties, the local averaged correlation function as introduced
in Eq.~(\ref{eq:summed_g2}) reaches the limit of $\mathcal{G}^{(2)}_{\mu \mu} \simeq 2(\sum_i n_i)^2 \equiv \frac12 \langle \hat N \rangle^2$,
exceeding the number fluctuations of a coherent state by a factor of two.

\begin{figure}[htb]
 \includegraphics{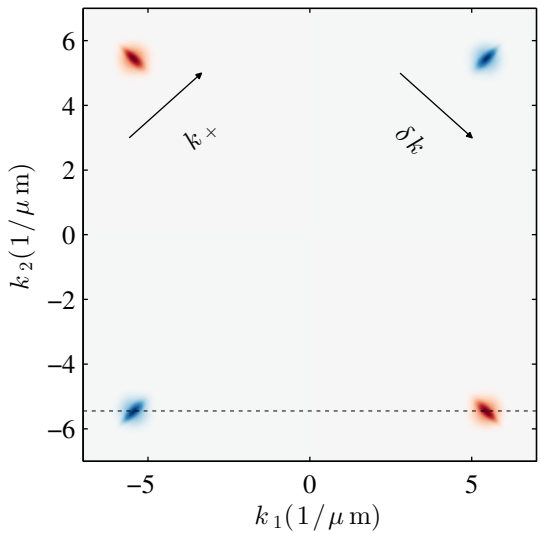}
  \caption{(color online)  False-color plot of non-normalized correlation function $G^{(2)}(k_1,k_2;t=1.2~\mathrm{ms})$.
  Contributions of $G^{(1)}$ and $M$ in Eq.~(\ref{g2_w}) are shown in blue and red hues, respectively.
  The dashed line indicates the position of the cut shown in Fig.~\ref{fig_den}(b).
  Arrows indicate the axes of Fig.~\ref{fig_avgd}.}
  \label{fig_k1k2}
 \end{figure}

\begin{figure}[htb!]
 \includegraphics{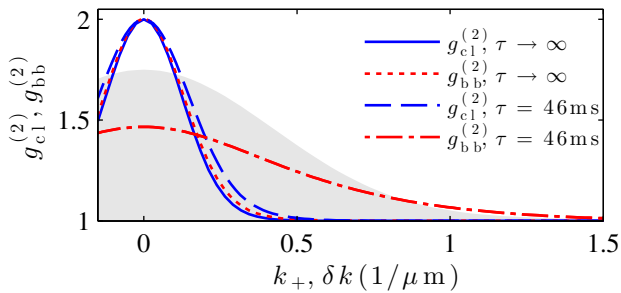}
  \caption{(color online)  Averaged, normalized second-order correlation functions $g^{(2)}(\delta k;t=1.2~\mathrm{ms})$, as obtained in experiments.
  Solid line: momentum-space peak near $k_1=k_2=k_0$, shown along the difference coordinate $\delta k = k_1 - k_2$ as indicated by the arrow in Fig.~\ref{fig_k1k2}.
  Dotted line: peak near $-k_1 = k_2 = k_0$, along the sum coordinate $k_+ = k_1 + k_2$.
  Dashed and dash-dotted lines: respective functions, taking into account the finite expansion time in time-of-flight momentum measurements.
  The grey shaded area is proportional to the normalization $\int \rho(k)\rho(k+\delta k) \mathrm{d}k$.}
 \label{fig_avgd}
\end{figure}

The typical tool to capture local second-order correlations in experiments are collinearly integrated functions of the type
\begin{eqnarray}
g_\mathrm{cl}^{(2)}(\delta \K;t) &=& \frac{\int G^{(2)}(\K,\K+\delta \K;t) \mathrm{d}\K}{ \int \rho(\K;t) \rho(\K+\delta\K;t) \mathrm{d}\K}
\end{eqnarray}
where the integrals run over an appropriately chosen momentum region~\cite{Perrin07,Kheruntsyan12}.
For symmetric, non-local correlations, back-to-back integration of the type
\begin{eqnarray}
g_\mathrm{bb}^{(2)}(\K_+;t) &=& \frac{\int G^{(2)}(\K,\K_+- \K;t) \mathrm{d}\K}{ \int \rho(\K;t) \rho(\K_+ - \K;t) \mathrm{d}\K}
\end{eqnarray}
is used.
The corresponding normalized functions for our system  $g_\mathrm{cl}^{(2)}(\delta_k;t), g_\mathrm{bb}^{(2)}(\delta_k;t)$ are shown in Fig.~\ref{fig_avgd} at $t=1.2~\mathrm{ms}$.
For both local (solid) and non-local (dotted) functions, correlations peaks, which do not span the entire populated range (grey area) and follow a Gaussian shape, are clearly present.

In the next step we take towards future comparison with experiments, we present the results not in momentum space, but rather using
real-space data calculated after some finite time $\tau$ of ballistic expansion. Only in the limit of $\tau \rightarrow \infty$ (far field),
the real-space data is equivalent to the initial momentum space distribution (if
the expanding clouds are sufficiently dilute, so that the mean-field repulsion can be safely neglected).
In~\cite{bucker}, the expansion time was $\tau = 46~\mathrm{ms}$, which was sufficient to resolve the twin-beam peaks. Nevertheless, the system was not fully in the far-field regime yet, which may have some
impact on the correlation functions.
As shown in Fig.~\ref{fig_avgd}, the finite expansion time affects the back-to-back peak at $(k_0,-k_0)$ much more strongly than the collinear HBT peak,
leading to smearing of the measured $g_\mathrm{bb}^{(2)}(k_+;t)$ (dash-dotted line) over the entire size of the twin-beam packets. This observation is consistent with some previous results \cite{wasak_raman}.
Intuitively, the broadening effect is related to the random position of scattering events along $x$ within the size of the initial cloud,
which is non-vanishing with respect to the expanded size of the twin-beam peaks.
On the other hand, the local correlation function $g_\mathrm{cl}^{(2)}(\delta k;t)$ (dashed line) remains largely unaffected.

Note that although at every instant of the evolution, the field operator $\hat\delta(\K,t)$ can be written as a sum of independently squeezed modes,
at very early times the division between the right and left modes is unjustified, because the two peaks are not yet fully separated.
However, at $t=0.1$\,ms when the density distribution is broad, the number of scattered atoms is $\langle\hat N\rangle\simeq2$.
Therefore, the system at such early time is hardly accessible experimentally so the quantum state of much less then a single particle is not of interest.
As soon as the two peaks are well-formed, at $t\approx0.3$\,ms, with $\langle\hat N\rangle\simeq8$ scattered atoms, all the general considerations from Sec.~\ref{general} apply.

\begin{figure}[htb]
  \includegraphics[clip, scale=0.35]{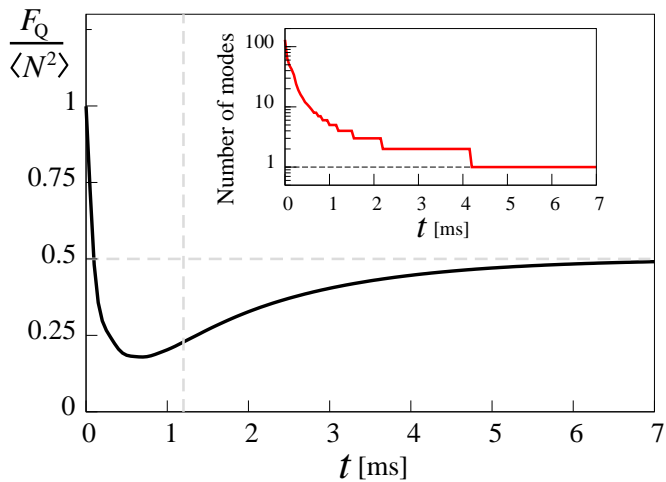}
  \caption{(color online) The Quantum Fisher Information as a function of time, normalized to $\langle\hat N^2\rangle$. The horizontal grey
    dashed line denotes the best possible value for the Bogoliubov system, which is achieved in a regime, where only a single pair of left/right modes is relevant.
    The vertical grey dashed line denotes the time
    $t=1.2$\,ms, when the Bogoliubov simulation should be interrupted. In the inset, we show the number of pairs of left/right modes that have at least 10\% of occupation of the largest modes.
    We see that only around 4.2\,ms, the two-mode approximation is valid, as denoted by the horizontal dashed line.
  }\label{fig_qfi}
\end{figure}

Finally, in Fig.~\ref{fig_qfi} we plot the QFI from Eq.~(\ref{qfi}) as a function of time and normalized to the Heisenberg Limit, i.e. $F_{\rm Q}\big/\langle\hat N^2\rangle$.
Instead of interrupting the simulation at $1.2$\,ms, where the scattered fraction of atoms becomes non-negligible and particle number conservation is strongly violated,
we extend the calculation up to $7$\,ms, when the number of scattered atoms significantly exceeds 15\% of $N_0$.
This is done solely to illustrate that, once the population of one of the modes dominates, $F_{\rm Q}\rightarrow\frac12\langle\hat N^2\rangle$, as argued in Sec.~\ref{entanglement}.
Note that the dominance of a single mode pair at long times is also predicted by the number-conserving theory \cite{bucker3}, justifying this proceeding.
Indeed, in the inset, we show the number of pairs of right/left modes which have an occupation bigger or equal to 10\% of the largest mode.
This approximately tells, how many modes are significantly occupied in the system.
At early times, there are over 100 pairs of modes. At $1.2$\,ms there are still 5 significantly occupied pairs, and only around 4.2\,ms a single pair of modes starts to dominate.
At the same time the QFI approaches its upper bound.

\section{Concluding remarks}

\label{conclusions}

We have developed a simple Bogoliubov model describing twin-atom beam experiments similar to Refs.~\cite{bucker,lattice}. Due to the elongated geometry of the trapping potential, the dynamics is one-dimensional.
As a consequence, the final step of our method can be easily solved numerically without the need for any stochastic method.
Furthermore, basic information about the scattered particles can directly be drawn from the general properties of the solution of the Bogoliubov equations. In this way, we can quantitatively characterize the mode structure and correlation functions of the scattered atoms.
Also, quite generally, we can show that the population imbalance between the two beams is ideally squeezed and that the system is strongly entangled.
These general observations can be applied to most recent experiments, where the atomic pairs scatter into two well-separated regions. Finally,
using the notion of the Quantum Fisher Information, we have derived a simple lower bound for the useful entanglement of the system. This expression employs only the average number of scattered particles and the number of occupied modes.

Having understood the fundamental properties of few-mode twin beams, further steps can be made to take into account more specific issues of experimental implementations. 
A general feature of strongly elongated Bose gases at realistic temperatures, such as the source cloud in \cite{bucker}, is the quasi-condensation~\cite{petrov}, 
where the coherence along the $x$-axis is limited due to thermal phase fluctuations.
Although these fluctuations do not alter the general considerations of Sec.~\ref{general}, they affect the emission dynamics \cite{bucker3},
and also might have influence on the spatial properties of both the density and the correlation functions \cite{wasak_raman}.
In future, our method could be applied to the analysis of some more complicated schemes, basing on the twin-beam setup, such as the Rarity-Tapster-type experiments \cite{rarity}. Also,
according to our results, the two-region state could be used as an input to the Mach-Zehnder-like interferometer, similarly to \cite{smerzi}.

Finally, note that the Bogoliubov approximation neglects the secondary collisions between the scattered particles and the atoms from the source cloud. 
When a scattered atom propagates through the BEC, the number of secondary collisions is bounded from above by $N_{\rm col}=16\pi a^2\,n\,R_{\rm tf}$, 
where $n$ is the peak density of the BEC. By plugging in the experimental numbers, we obtain $N_{\rm col}=0.38$, which well justifies the use of the Bogoliubov approximation.

\section{Acknowledgements}
J. Ch. acknowledges the Foundation for Polish Science International TEAM Programme co-financed by the EU European Regional Development Fund.
T.W. and P. Sz. acknowledge the Foundation for Polish Science International Ph.D. Projects Programme co-financed by the EU European Regional Development Fund.
R.B. acknowledges support from the Austrian Science Fund (FWF) projects CAP (1607-N16), and Atom Chip (Z118-N16), and the FWF doctoral programme CoQuS (W1210).
This research was supported by the National Science Center grants no. DEC-2011/03/D/ST2/00200 and N202 167840.


\begin{thebibliography}{40}

\bibitem{kwiat} Paul G. Kwiat, Klaus Mattle, Harald Weinfurter, Anton Zeilinger, Alexander V. Sergienko, and Yanhua Shih, Phys. Rev. Lett. {\bf 75}, 4337 (1995)
\bibitem{horodecki} R. Horodecki, P. Horodecki, M. Horodecki, and K. Horodecki, Rev. Mod. Phys. {\bf 81}, 865 (2009)

\bibitem{Reid09} M.D.~Reid, P.D.~Drummond, W.P.~Bowen, E.G.~Cavalcanti, P.H.~Lam, H.A.~Bachor, U.~L.~Andersen, G.~Leuchs, Rev.~Mod.~Phys.~\textbf{81}, 1727 (2009).

\bibitem{wootters1} Charles H. Bennett, Gilles Brassard, Claude Cr\'epeau, Richard Jozsa, Asher Peres, and William K. Wootters, Phys. Rev. Lett. {\bf 70}, 1895 (1993)
\bibitem{wootters2} Charles H. Bennett, Gilles Brassard, Sandu Popescu, Benjamin Schumacher, John A. Smolin, and William K. Wootters, Phys. Rev. Lett. {\bf 76}, 722 (1996)

\bibitem{giovanetti} {V. Giovanetti, S. Lloyd and L. Maccone, Science {\bf 306}, 1330 (2004)}
\bibitem{pezze}{L. Pezz\'e and A. Smerzi, Phys. Rev. Lett. {\bf 102}, 100401 (2009)}


\bibitem{esteve}{J. Est\'eve, C. Gross, A. Weller, S. Giovanazzi and M. K. Oberthaler, Nature {\bf 455}, 1216 (2008)}
\bibitem{maus}{Kenneth Maussang, G. Edward Marti1, Tobias Schneider, Philipp Treutlein, Yun Li, Alice Sinatra, Romain Long, J\'er\^ome Est\`eve, and Jakob Reichel, Phys. Rev. Lett. {\bf 105}, 080403 (2010)}
\bibitem{berrada}{T. Berrada, S. van Frank, R. B\"ucker, T. Schumm, J.-F. Schaff, and J. Schmiedmayer, Nat. Comm. {\bf 4}, 2077 (2013)}

\bibitem{sor}{A. Sorensen, L.-M. Duan, J.I. Cirac and P. Zoller, Nature {\bf 69}, 63 (2001)}

\bibitem{riedel} {Max F. Riedel,  Pascal B\"ohi,  Yun Li,  Theodor W. H\"ansch,  Alice Sinatra and  Philipp Treutlein, Nature {\bf 464}, 1170 (2010)}
\bibitem{gross3}{C. Gross, T. Zibold, E. Nicklas, J. Esteve and M. K. Oberthaler, Nature {\bf 464}, 1165 (2010)}
\bibitem{app}{J. Appel, P. J. Windpassinger, D. Oblak, U. B. Hoff, N. Kj\ae rgaard, and E. S. Polzik, PNAS {\bf 106}, 10960 (2009)}

\bibitem{leroux}{Ian D. Leroux, Monika H. Schleier-Smith, and Vladan Vuleti\'c, Phys. Rev. Lett. {\bf 104}, 073602 (2010)}
\bibitem{schl}{Monika H. Schleier-Smith, Ian D. Leroux, and Vladan Vuleti\'c, Phys. Rev. Lett. {\bf 104}, 073604 (2010)}

\bibitem{Perrin07} A.~Perrin, H.~Chang, V.~Krachmalnicoff, M.~Schellekens, D.~Boiron, A.~Aspect, C.I.~Westbrook, \PRL{99}, 150405 (2007).
\bibitem{Krachmalnicoff10} V.~Krachmalnicoff, J.-C.~Jaskula, M.~Bonneau, V.~Leung, G.~B.~Partridge, D.~Boiron, C.~I.~Westbrook,
  P.~Deuar, P.~Zi\'n, M.~Trippenbach, K.~V.~Kheruntsyan, \PRL{104}, 150402 (2010).
\bibitem{Jaskula10} J.-C.~Jaskula, M.~Bonneau, G.~B.~Partridge, V.~Krachmalnicoff, P.~Deuar, K.~V.~Kheruntsyan, A.~Aspect, D.~Boiron, C.~I.~Westbrook, \PRL{105}, 190402 (2010)
\bibitem{Kheruntsyan12} K.V.~Kheruntsyan, J.-C.~Jaskula, P.~Deuar, M.~Bonneau, G.B.~Partridge, J.~Ruaudel, R.~Lopes, D.~Boiron, C.I.~Westbrook, \PRL{108}, 260401 (2012).

\bibitem{dall} R. G. Dall, L. J. Byron, A. G. Truscott, G. R. Dennis, M. T. Johnsson, and J. J. Hope, Phys. Rev. A {\bf 79}, 011601 (2009)
\bibitem{pertot} D. Pertot, B. Gadway, and D. Schneble, Phys. Rev. Lett. {\bf 104}, 200402 (2010)
\bibitem{rugway} Wu RuGway, S. S. Hodgman, R. G. Dall, M. T. Johnsson, and A. G. Truscott, Phys. Rev. Lett. {\bf 107}, 075301 (2011)


\bibitem{Vogels02} J.M.~Vogels, K.~Xu, W.~Ketterle, \PRL{89}, 020401 (2002).
\bibitem{deng}{L. Deng,  E. W. Hagley,  J. Wen,  M. Trippenbach,  Y. Band,  P. S. Julienne,  J. E. Simsarian,  K. Helmerson,  S. L. Rolston and  W. D. Phillips, Nature {\bf 398}, 218 (1999)}

\bibitem{campbell}{Gretchen K. Campbell, Jongchul Mun, Micah Boyd, Erik W. Streed, Wolfgang Ketterle, and David E. Pritchard, Phys. Rev. Lett. {\bf 96}, 020406 (2006)}
\bibitem{lattice} M. Bonneau, J. Ruaudel, R. Lopes, J.-C. Jaskula, A. Aspect, D. Boiron, and C. I. Westbrook, Phys. Rev. A {\bf 87}, 061603 (2013)

\bibitem{bucker} R. B\"ucker, J. Grond, S. Manz, T. Berrada, T. Betz, C. Koller, U. Hohenester, T. Schumm, A. Perrin and J. Schmiedmayer, Nat. Phys. {\bf 7}, 608 (2011)
\bibitem{smerzi} B. L\"ucke,  M. Scherer,  J. Kruse,  L. Pezz\'e, F. Deuretzbacher, P. Hyllus, O. Topic, J. Peise, W. Ertmer, J. Arlt, L. Santos, A. Smerzi and C. Klempt,
  Science {\bf 11}, 773 (2011)

\bibitem{gross2}{C. Gross,  H. Strobel,  E. Nicklas,  T. Zibold,  N. Bar-Gill,  G. Kurizki and  M. K. Oberthaler, Nature {\bf 480}, 219 (2011)}
\bibitem{bookjans}{Eva M. Bookjans, Christopher D. Hamley, and Michael S. Chapman, Phys. Rev. Lett. {\bf 107}, 210406 (2011)}

\bibitem{braun2}  S. L. Braunstein, Phys. Rev. A {\bf 71}, 055801 (2005).
\bibitem{sacha} J. Dziarmaga and K. Sacha, J. Phys. B {\bf 39}, 57 (2006)
\bibitem{single} J. Chwede\'nczuk, P. Zi\'n, K. Rz\c{a}\.zewski, and M. Trippenbach, Phys. Rev. Lett. {\bf 97}, 170404 (2006)






\bibitem{ketterle} J. M. Vogels, K. Xu, and W. Ketterle, Phys. Rev. Lett. {\bf 89}, 020401 (2002)
\bibitem{perrin} A. Perrin, H. Chang, V. Krachmalnicoff, M. Schellekens, D. Boiron, A. Aspect, and C. I. Westbrook, Phys. Rev. Lett. {\bf 99}, 150405 (2007)

\bibitem{squeezing} J.-C. Jaskula, M. Bonneau, G. B. Partridge, V. Krachmalnicoff, P. Deuar, K. V. Kheruntsyan, A. Aspect, D. Boiron, and C. I. Westbrook, Phys. Rev. Lett. {\bf 105}, 190402 (2010)

\bibitem{wasak} P. Deuar, T. Wasak, P. Zi\'n, J. Chwede\'nczuk and M. Trippenbach, Phys. Rev. A {\bf 88}, 013617 (2013)

\bibitem{marino} A. M. Marino, V. Boyer, and P. D. Lett, Phys. Rev. Lett. {\bf 100}, 233601 (2008)

\bibitem{helstrom} Carl W. Helstrom, {\it Quantum Detection and Estimation Theory}, Academic Press (1976)
\bibitem{cramer} H. Cram{\'e}r, \emph{Mathematical Methods of Statistics}, (Princeton Univ. Press, Princeton, NJ, 1946).


\bibitem{braun} S. L. Braunstein and C. M. Caves, Phys. Rev. Lett. {\bf 72}, 3439 (1994).

\bibitem{sorensen} A. S{\o}rensen, L.-M. Duan, J. I. Cirac and P. Zoller, Nature {\bf 409}, 63 (2001)
\bibitem{hyllus} P. Hyllus, L. Pezz\'e, and A. Smerzi, Phys. Rev. Lett. {\bf 105}, 120501 (2010)

\bibitem{szan} P. Sza\'nkowski, T. Wasak, J. Chwede\'nczuk and M. Trippenbach, in preparation

\bibitem{bucker2} R. B\"ucker, T. Berrada, S. van Frank, J.-F. Schaff, T. Schumm, J. Schmiedmayer, G. J\"ager, J. Grond and U. Hohenester, J. Phys. B {\bf 46}, 104012 (2013)
\bibitem{petrov} D.S. Petrov, G.V. Shlyapnikov and J.T.M. Walraven, Phys. Rev. Lett. {\bf 87}, 050404 (2001)
\bibitem{dalfovo} F. Dalfovo, S. Giorgini, L. P. Pitaevskii and S. Stringari, Rev. Mod. Phys. {\bf 71}, 463 (1999)

\bibitem{bucker3}{R. B\"ucker, U. Hohenester, T. Berrada, S. van Frank, A. Perrin, S. Manz, T. Betz, J. Grond, T. Schumm, and J. Schmiedmayer , Phys. Rev. A {\bf 86}, 013638 (2012)}


\bibitem{wasak_raman} T. Wasak, J. Chwede\'nczuk, P. Zi\'n, and M. Trippenbach, Phys. Rev. A {\bf 86}, 043621 (2012)



\bibitem{rarity} J. G. Rarity and P. R. Tapster, Phys. Rev. Lett. {\bf 64}, 2495 (1990)

%\bibitem{gross}{Ch. Gross, J. Phys. B {\bf 10}, 103001 (2012)}

\end{thebibliography}
\end{document}